\documentstyle[12pt,psfig,aaspp4]{article}	
\def\mathmode#1{\ifmmode {#1} \else {$#1\mkern-5mu$} \fi} \def\aa{A\&A}
 \def\aj{AJ} \def\apj{ApJ} \def\apjs{ApJS}
\def\mnras{MNRAS}  \def\hst{{\it HST\/}}

 \def\V6{$V_{606}$} \def\V5{$V_{555}$}
\def\I8{$I_{814}$}  
      \def\ie{{\it i.e.\/}} 
  \def\HST{{\it HST\/}}

\begin{document}

\title{HST Observations of Galactic Globular Cluster Cores. II. 
NGC~6273 and the Problem of Horizontal-Branch Gaps.\altaffilmark{1}}
\author{G. Piotto\altaffilmark{2}, M. Zoccali\altaffilmark{2},
I.~R. King\altaffilmark{3}, S.\ G.\ Djorgovski\altaffilmark{4},
C. Sosin\altaffilmark{5},
R.\ M. Rich\altaffilmark{6}, and G.\ Meylan\altaffilmark{7}}

\altaffiltext{1}{Based on observations with the NASA/ESA {\it Hubble Space 
Telescope}, obtained at the Space Telescope Science Institute, which is 
operated by AURA, Inc., under NASA contract NAS5-26555.}

\altaffiltext{2}{Dipartimento di Astronomia, Universit\` a di Padova,
Vicolo dell'Osservatorio 5, I-35122 Padova, Italy; piotto@pd.astro.it,
zoccali@pd.astro.it}

\altaffiltext{3}{Astronomy Department, University of California,
Berkeley, CA 94720-3411; king@glob.berkeley.edu}

\altaffiltext{4}{California Institute of Technology, MS 105-24,
Pasadena, CA 91125; george@deimos.caltech.edu}

\altaffiltext{5}{Astronomy Department, University of California,
Berkeley, CA 94720-3411; csosin@sirius.com}

\altaffiltext{6}{Department of Physics and Astronomy,
Division of Astronomy and Astrophysics, University of California,
Los Angeles, CA 90095-1562; rmr@astro.ucla.edu}

\altaffiltext{7}{European Southern Observatory,
Karl-Schwarzschild-Strasse 2, D-85748 Garching bei M\"unchen, Germany;
gmeylan@eso.org}


\begin{abstract}

We present observations of the center of the Galactic globular cluster
NGC~6273, obtained with the HST/WFPC2 camera as part of the snapshot
program GO-7470. A $B$, $V$ color--magnitude diagram (CMD) for
$\sim$28,000 stars is presented and discussed.  The most prominent
feature of the CMD, identified for the first time in this paper, is
the extended horizontal branch blue tail (EBT) with a clear
double-peaked distribution and a significant gap.  The EBT of NGC~6273
is compared with the EBTs of seven other globular clusters for which we
have a CMD in the same photometric system. From this comparison we
conclude that all the globular clusters in our sample with an EBT show
at least one gap along the HB, which could have similar origins.  A
comparison with theoretical models suggests that at least some of
these gaps may be occuring at a particular value of the stellar mass,
common to a number of different clusters. From the CMD of NGC 6273 we obtain a
distance modulus $(m-M)_V=16.27\pm0.20$. We also estimate an average
reddening E($B-V$)=0.47$\pm$0.03, though the CMD is strongly affected
by differential reddening, with the relative reddening spanning a
$\Delta E(B-V)\sim 0.2$ magnitude in the WFPC2 field.  A luminosity
function for the evolved stars in NGC~6273 is also presented and
compared with the most recent evolutionary models.

\end{abstract}



\section{Introduction}\label{intro}

A growing number of Galactic globular clusters (GGC) is found to show
extended-horizontal-branch blue tails (EBT), which sometimes reach the
He-burning main sequence, indicating that some of the stars must have
lost (almost) all of their envelope during the red giant branch (RGB)
phase.  The EBTs represent a puzzle in the stellar evolution models.
In many GGCs there is another peculiarity:\ a discontinuity in the
stellar distribution along the EBT, which sometimes results in a sort
of gap, i.e., a region clearly underpopulated in stars (Sosin et al.\
1997a, Catelan et al.\ 1998). No clear explanation for the origin of
the gaps is available at present.  These abnormalities in the HB
probably represent one of the most extreme of the mixed bag of
anomalies that are sometimes lumped under the term ``second-parameter
problem'' (cf.\ Ferraro et al.\ 1998).

In this paper, we present a new cluster with both an EBT and a
significant gap on it: NGC~6273. This object is one of the 46 clusters
we are hoping to observe within our ongoing \HST\ snapshot programs.
NGC~6273 (M19) ($\alpha_{1950}=16^{{\rm h}}59^{{\rm m}},
\delta_{1950}=-26^\circ11'$) is a medium-concentration ($c=1.5$,
Djorgovski 1993), intermediate-metallicity ([Fe/H] $=-1.68$, Zinn \&
West 1984) cluster, located toward the direction of the Galactic bulge
($l=357^\circ$, $b=+9^\circ$).  NGC~6273 is the second most luminous
cluster, after $\omega$ Cen, in the Djorgovski (1993) compilation. In
a photographic image NGC~6273 looks very similar to $\omega$ Cen.  The
other noteworthy property of NGC~6273 is its high ellipticity,
$\epsilon=0.28$, which makes M19 the most flattened of the GGC in the
White \& Shawl (1987) catalogue, and probably in the entire Galaxy.  The
only available color--magnitude diagram (CMD) is by Harris, Racine, \&
De Roux (1976), and it barely reaches the HB level. The CMD looks
quite dispersed due to field contamination and differential reddening,
estimated by the same authors to be around 0.2 magnitude in $B-V$.
Here we present a new CMD, extending to $\sim$2 magnitudes below the
turnoff (TO), and corrected for differential reddening..


\section{Observations and Analysis}\label{obs}

The center of the GGC NGC~6273 was observed with the \hst\ WFPC2
camera through the filters F439W and F555W ($=$ \hst\ $B,V).$ A series
of short and long exposures was secured in order to cover 
a large magnitude range (see Table 1 for a detailed log of the
observations).  The longer F439W exposures were repeated in order to
allow removal of cosmic rays. In both cases, the cluster center was
placed in the PC field.

\begin{deluxetable}{cccc}
\tablewidth{24pc}
\tablenum{1}
\tablecaption{Log of the observations \label{logobs}}
\tablehead{
\colhead{Object} &
\colhead{Filter} &
\colhead{Exp. time} &
\colhead{Date}}
\startdata
NGC~6273 &  F555W  &   10s	&   May 5, 1998   \\
   ''    &  F555W  &   40s	&  	''	  \\
   ''    &  F439W  &   40s	&  	''	  \\
   ''    &  F439W  &  160s	&  	''	  \\
   ''    &  F439W  &  160s	&  	''	  \\
\tableline
\enddata
\end{deluxetable}

The images were pre-processed by the standard \hst\ pipeline.
Following the procedures described by Silbermann et al.\ (1996), we
masked out the vignetted pixels and bad pixels and columns using a
vignetting mask created by P.\ B.\ Stetson, together with the
appropriate data-quality file for each frame. We have also multiplied
each frame by a pixel area map (also provided by P.\ B.\ Stetson) in
order to correct for the geometric distortion (Silbermann et al.\
1996).  The saturated pixels of the brightest stars in the longest
exposures were corrected as in Cool, Piotto, \& King (1996), using the
shorter exposures.

The photometric reduction was carried out following the same procedure
that is described in detail in Piotto et al.\ (1999a) for a similar set
of images acquired within the same HST program. The raw instrumental
magnitudes were transformed into the standard Johnson $BV$ system
using Eq.\ (8) in Holtzman et al.\ (1995) and the coefficents in their
Table 7.

As this paper makes extensive use of star counts, particular attention
was devoted to estimating the completeness in all the principal branches
of the CMD.  The completeness corrections were determined by
artificial-star experiments, again as described in Piotto et al.\ (1999a).


\section{Color--Magnitude Diagrams}\label{cmds}

\begin{figure}
\psfig{figure=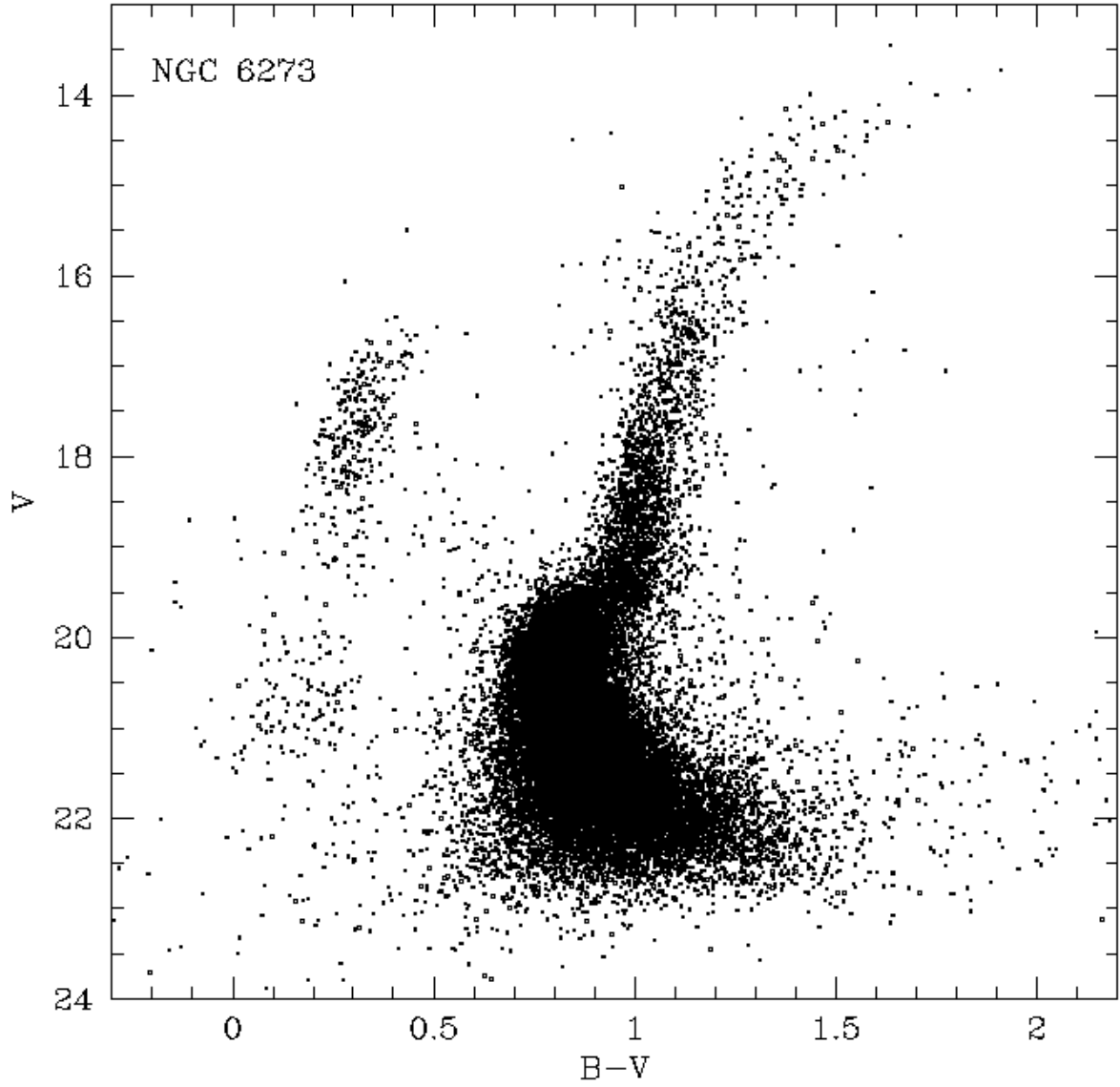,width=16cm}
\caption[cmdor.ps]{Color--magnitude diagram for $\sim 28,000$ stars in the 
central region of NGC~6273. All the stars identified in the PC camera and in
the three WF cameras are shown. 
\label{cmdor}}
\end{figure}

The CMD for $\sim$28,000 stars in NGC~6273 is shown in Fig.\
\ref{cmdor}.  The diagram extends from almost the tip of the giant
branch (stars brighter than $V \approx 15$ are saturated even in the
shortest-exposure frames) to $\sim$2 magnitudes below the TO. All
the relevant branches of the CMD are well defined. As in all the CMDs
of the GGC cores, a well defined sequence of blue stragglers is clearly 
seen, extending from the TO up to the HB.  The most
prominent features in Fig.\ \ref{cmdor} are the
extended blue HB and the striking gap centered at $V\sim19.5$.  
Neither of these features was reached by the previously
published CMD.

\subsection{Reddening}\label{red}

\begin{figure}
\psfig{figure=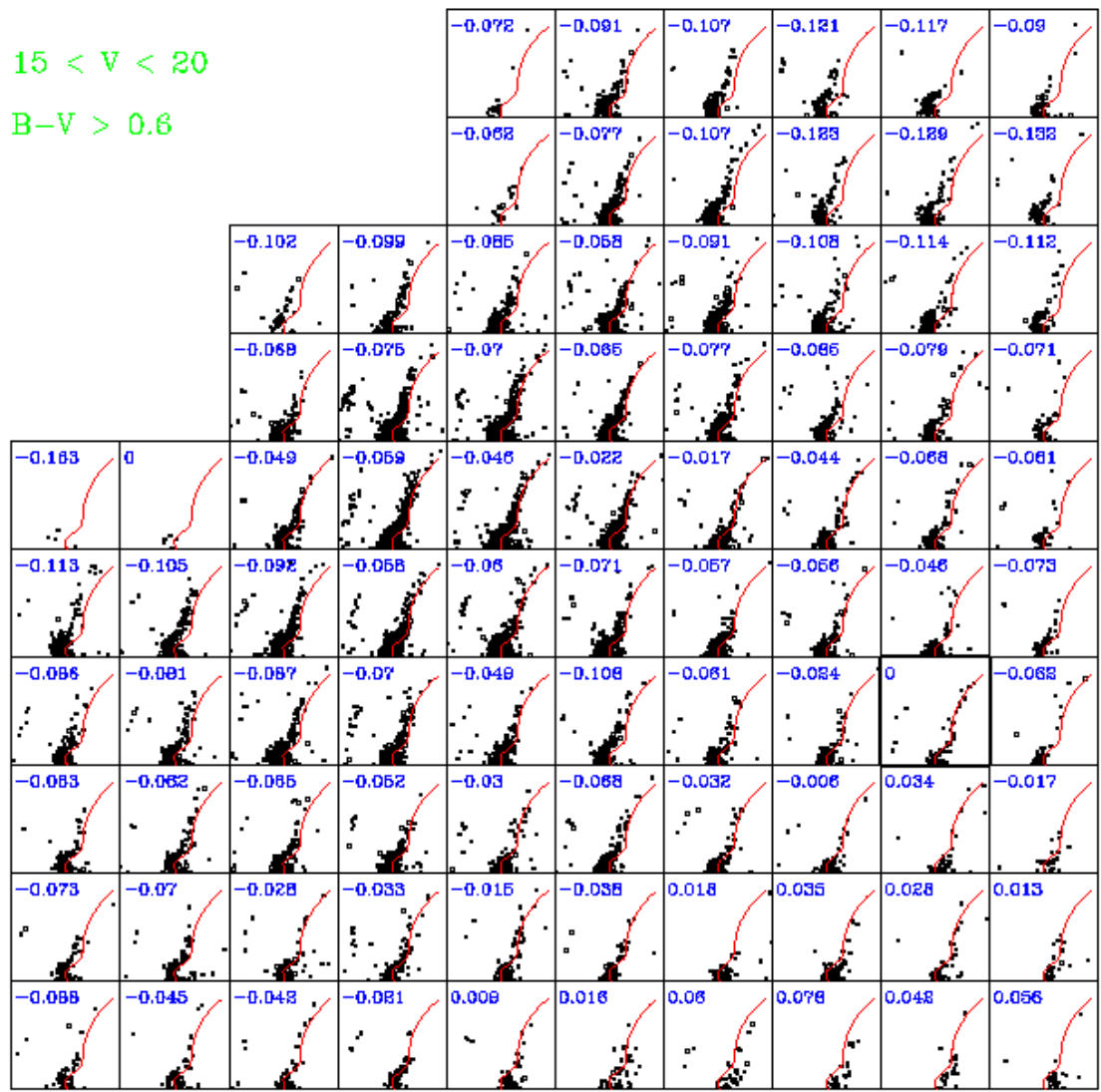,width=16cm}
\caption[rgrid.ps]{The WFPC2 field has been divided into 86
$15\times15$ arcsec$^2$ regions in order to study the spatial
variation of the reddening over the central region of NGC~6273. 
The plot reproduces the field covered by the WFPC2, with the PC at
the top left. The CMD of the region (9,4) has been arbitrarily 
adopted as reference. The reddening relative to the reference CMD 
is displayed at the top left of each region.
\label{rgrid}}
\end{figure}

All the sequences in Fig.\ \ref{cmdor} are broadened. The broadening
is much larger than the typical photometric errors, at all magnitudes.
According to the artificial-star experiments, the dispersion in color
due to photometric errors along the RGB (for $17<V<19$) is $\Delta
(B-V)=0.03$ (standard deviation around the mean value).  In view of
the position of NGC~6273 ($l=357^\circ$, $b=+9^\circ$), we can expect
some differential reddening. In order to test this hypothesis, we
divided our field into 86 $15\times15$ arcsec$^2$ regions (Fig.\
\ref{rgrid}). The dimensions of the regions are a compromise between
the need to have the highest spatial resolution and the need to have a
sufficient number of stars to be able to identify the giant-branch
position.  For each sub-image we obtained a CMD. We selected one of
these CMDs as the reference CMD, and extracted its fiducial points by
drawing a line by hand. In Fig.\ \ref{rgrid} the CMDs for each region
are compared with the fiducial line obtained by fitting a spline to
the fiducial points. The effects of the differential reddening are
clearly visible. For each region the reddening relative to the
fiducial sequence is indicated. The relative reddening has been
calculated as follows.  In each spatial region, for each star with
$15<V<20$ and $(B-V)>0.6$, we calculated its distance from the
fiducial CMD along the reddening line [defined by the relation
$V=3.2(B-V)$].  Each one of these distances is the resultant of two
components:\ $E(B-V)$ on the abscissa and $A_V$ on the ordinate. The
relative reddening is the median value of the abscissas of all the
distances.  Fig.\ \ref{rgrid} shows a radial trend of the relative
reddening, with the central region less reddened. The relative
reddening spans a $\Delta E(B-V)\sim 0.2$ magnitude in the WFPC2
field, confirming the results by Harris et al.\ (1976).  The standard
deviation around the average reddening [the mode of the reddening
distribution for the WFPC2 field is $E(B-V)=0.41$] is 0.05 magnitude.
We also determined the average reddening of NGC~6273 by comparing the
location of its EBT with the EBTs of two other clusters:\ NGC~1904
(Sosin et al.\ 1997b, Piotto et al.\ 1999b) and NGC~6205 (M13, Zoccali
\& Piotto 1999).  The turning down of the blue part of the HB is due
to the saturation of the $B-V$ color as the bolometric correction
increases sharply at $T_{\rm eff}\sim 10,000$ K. It is thus not very
sensitive to metallicity or age.  We selected these two clusters for
the comparison because they show the same peculiar HB as NGC~6273.
Besides, in Zinn \& West (1984) the three clusters have the same
metallicity (formally within 0.03 dex). The data of NGC~1904 and
NGC~6205 are in the same HST photometric system as NGC~6273. By
comparing the HBs, we find that NGC~6273 is $0.45\pm0.02$ magnitude
redder than NGC~1904 and $0.47\pm0.02$ magnitude redder than
NGC~6205. Assuming $E(B-V)=0.01$ for NGC~1904 (Stetson \& Harris 1977)
and $E(B-V)=0.02$ for NGC~6205 (Djorgovski 1993), we have a mean
reddening for the reference fiducial points of Fig.\ \ref{rgrid}
$E(B-V)=0.47\pm0.03$ for our field in NGC~6273.

\begin{figure}
\psfig{figure=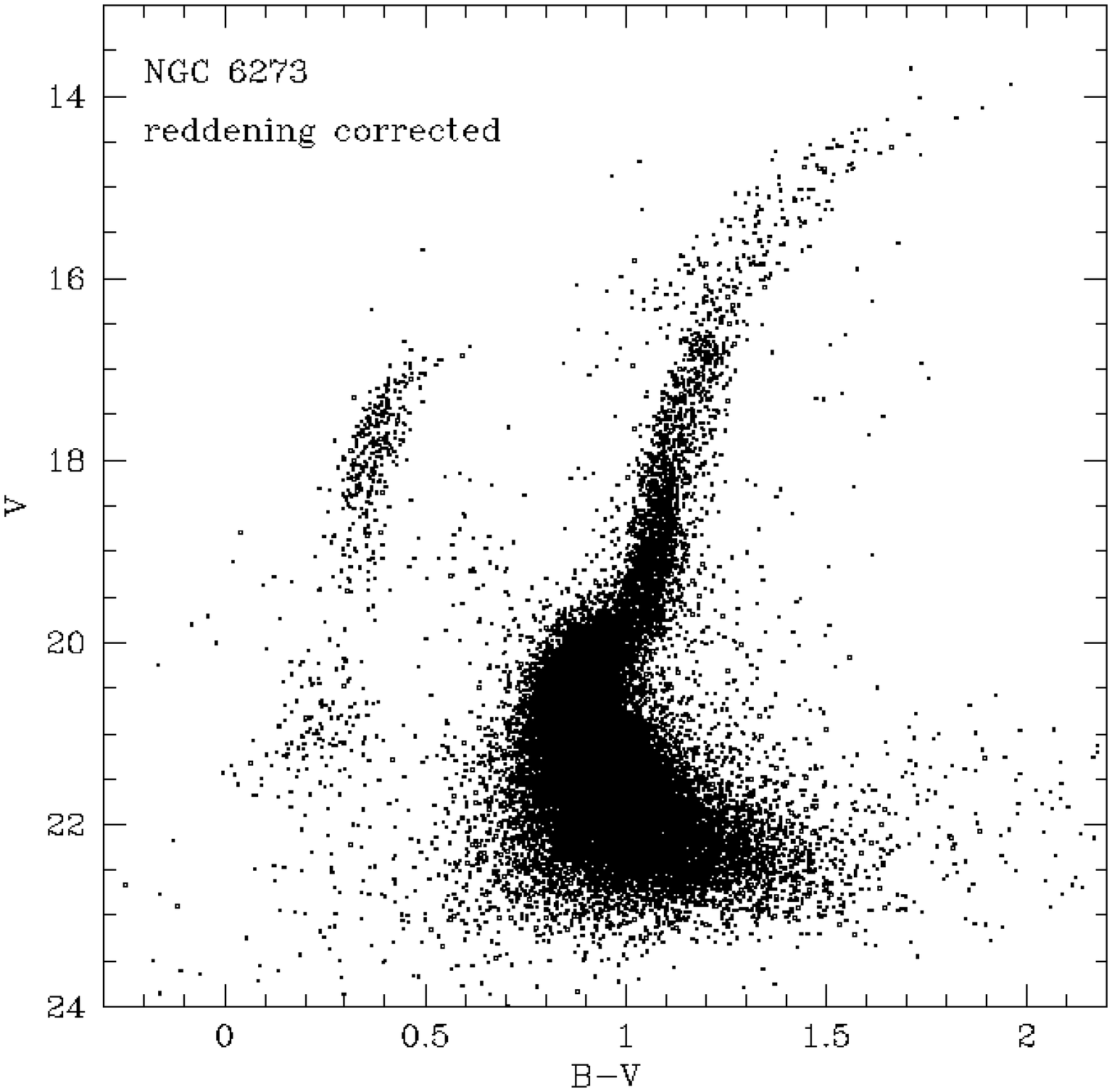,width=16cm}
\caption[cmdco.ps]{As in Fig.\ \ref{cmdor}, but after the relative reddenings
as determined in Fig.\ \ref{rgrid} have been removed. The broadness of
all the CMD sequences is still larger than expected from the
photometric errors, probably due to a residual differential reddening
acting on scales smaller than $15\times15$ arcsec$^2$.
\label{cmdco}}
\end{figure}

The reddening estimates in Fig.\ \ref{rgrid} can be used to
(partially) correct the CMD of Fig.\ \ref{cmdor} for the effects of
differential reddening. To the CMDs in each of the regions shown in
Fig.\ \ref{rgrid} we applied the relative reddening values indicated, 
and corrected the color and the magnitude of each star for reddening 
and the corresponding absorption $A_V=3.2E(B-V)$.  The
resulting CMD for the entire WFPC2 field is shown in Fig.\ \ref{cmdco}.
All the sequences in the CMD of Fig.\ \ref{cmdco} are sensibly narrower
than in Fig.\ \ref{cmdor}, but still broader than expected from the
artificial-star experiments. There are two possible explanations: {\it
i)} there might be some differential reddening on spatial scales smaller
than $15\times15$ arcsec$^2$; {\it ii)} there might be some dispersion
in the metal content of the stars in NGC~6273. Though the first
hypothesis seems the more likely, nevertheless in view of the similarity
of NGC~6273 to $\omega$ Cen discussed in Section~ \ref{intro}, a study
of the metallicity of its stars would be worth doing. 
The spread in the combined equivalent width of the lines in the Ca II 
triplet ($\Sigma Ca$) in Fig.\ 7 of Rutledge et al.\ (1997) might be 
an indication of a possible metallicity dispersion among the NGC~6273 
stars, but the number of measured stars is too small to allow any conclusion.

\subsection{Distance}\label{dist}

NGC~6273 has a blue HB without a horizontal portion, making the
distance determination quite uncertain.  So far, only three RR~Lyrae
have been identified in the direction of NGC~6273 (Clement and Sawyer
Hogg 1978).  Two of them fall within the RR~Lyrae instability strip of
NGC~6273, indicating possible cluster membership. The third one is 1.5
mag.  brighter than the HB, and Clement and Sawyer Hogg (1978)
consider it to be a non-member.

A more accurate distance estimate comes again from a comparison with
NGC~1904 and NGC~6205. In particular, the large statistical sample of
stars in the CMD of NGC~6273 and of the reference clusters allows us
to derive an accurate luminosity function (LF), where the red-giant
bump (Iben 1968) can be easily identified at $V_{\rm
bump}=16.80\pm0.10$.  In view of the similarity in metallicity (and
the similarity of the CMDs) we expect the bump to be located at the
same absolute magnitude in the three clusters.  From the LFs of
NGC~1904 and NGC~6205 (Zoccali \& Piotto 1999) we measured $V_{\rm
bump}=16.00\pm0.10$ for NGC~1904 and $V_{\rm bump}=14.85\pm0.10$ for
NGC~6205.  Adopting an apparent distance modulus ($m-M$)$_V=15.45$
(Ferraro et al.\ 1992) for NGC~1904 and ($m-M$)$_V=14.35$ (Ferraro et
al.\ 1997) for NGC~6205, we obtain for NGC~6273 ($m-M$)$_V=16.25$ and
($m-M$)$_V=16.30$, respectively.  From here on, we will adopt for
NGC~6273 ($m-M$)$_V=16.27\pm0.20$.  As there is no error associated with
the distance determination of NGC~1904 and NGC~6205, we have assumed a
typical error of 0.2 magnitude, though the method would allow
relative distance determination with sensibly smaller uncertainties.

Assuming $E(B-V)=0.47$, and adopting $A_V=3.2E(B-V)$, the absolute
distance modulus of NGC~6273 is $(m-M)_0=14.77\pm0.20$, i.e.,
NGC~6273 is at $d_\odot=9.0$ kpc from the Sun, at $Z=1.4$ kpc from the
Galactic plane.


\section{The distribution along the HB}\label{HB}

Certainly, the most prominent features of the CMDs of both Fig.\
\ref{cmdor} and Fig.\ \ref{cmdco} are the extended blue tail of the HB
and the gap along it.  As discussed in Section~\ref{intro}, there are
other clusters which have HBs apparently similar to that of NGC~6273.
One question that is still open is whether these features originate from
the same (though still unknown) physical process.  Certainly, the lack
of clear observational inputs makes more difficult the physical
interpretation of both the EBT and the gaps.  There are two problems
which complicate the analysis of the presently available data:\\ {\it
i)} the heterogeneity of the data gathered so far, in particular the
different photometric bands used;\\ {\it ii)} the shape of the HB,
particularly complicated in the $V$, $B-V$ plane because of the
saturation of the $B-V$ color as a temperature index when $B-V<0$.

In order to try to shed some light on this problem, we have collected
from the HST archive all the public WFPC2 images of GGC cores observed
with the same filters (F439W and F555W) that we used in our previous
HST GO-6095 program, and that we are still using in the presently
running GO-7470 and will use in GO-8118. At the present time, we have
CMDs for 29 GGCs in this homogeneous data set. Among these, at least
eight GGCs show blue tails with more or less pronounced gaps.

A detailed comparison of the positions of peaks and gaps as a function
of stellar mass or temperature for different theoretical models is
beyond the scope of the present paper, and will be addressed
elsewhere.  On a more empirical basis, as originally suggested by Rood
\& Crocker (1985, 1989), in order to analyze the distribution of the
HB stars it is useful to define a new coordinate, $l_{\rm HB}$, which
is linear along the HB ridge line. This coordinate removes the
saturation of the $(B-V)$ color as a function of the temperature.
Unfortunately, different authors (Ferraro et al.\ 1992, Dixon et al.\
1996, Catelan et al.\ 1998) have given different definitions of
$l_{\rm HB}$, besides the fact that sometimes the definition itself
has not been clear nor the measurement of $l_{\rm HB}$ easily
reproducible.

In order to linearize {\it all} the HBs in our data base, which is the
largest one presently available, we were led to define an $l_{\rm HB}$
which is itself slightly different from the previous ones. However, in
order to avoid confusion, and in the attempt to introduce a
``standard'' $l_{\rm HB}$, we describe in detail the recipe we used.

\begin{itemize}

\item First of all, the observed CMDs were de-reddened, as described in 
Section~\ref{red} for NGC~6273, by comparison with NGC~6205, adopting
an $E(B-V)=0.02$ for the latter.

\item The HB colors and magnitudes were then transformed according to
the following relations:
\[
c = 204.8 (B-V)_0 + 102.4\qquad\hbox{and}\qquad b = -42.67 M_V + 281.6.
\]

These new ``color'' and ``brightness'' coordinates allow mapping the CMD
onto a plane where 1 unit in the abscissa has the same length (in
centimeters) as 1 unit in the ordinate, mantaining the same scale as in
Ferraro et al.\ (1992) and Fusi Pecci et al.\ (1993).  This last
condition (which is at the basis of the choice of the numerical factors
in the above equation) is not strictly necessary for the comparison
among our clusters, but we preferred to keep our $l_{\rm HB}$ as close
as possible to the previous definitions.

\item The HB ridge lines (HBRL) were obtained by a spline interpolation of
hand-drawn fiducial points.

\item As shown in Fig.\ \ref{pro}, the projection of each HB star on the 
HBRL was determined as the point ($c_R$, $b_R$) that minimizes the
distance
\[
\sqrt{(c-c_R)^2 + (b-b_R)^2}.
\]
	
\begin{figure} 
\psfig{figure=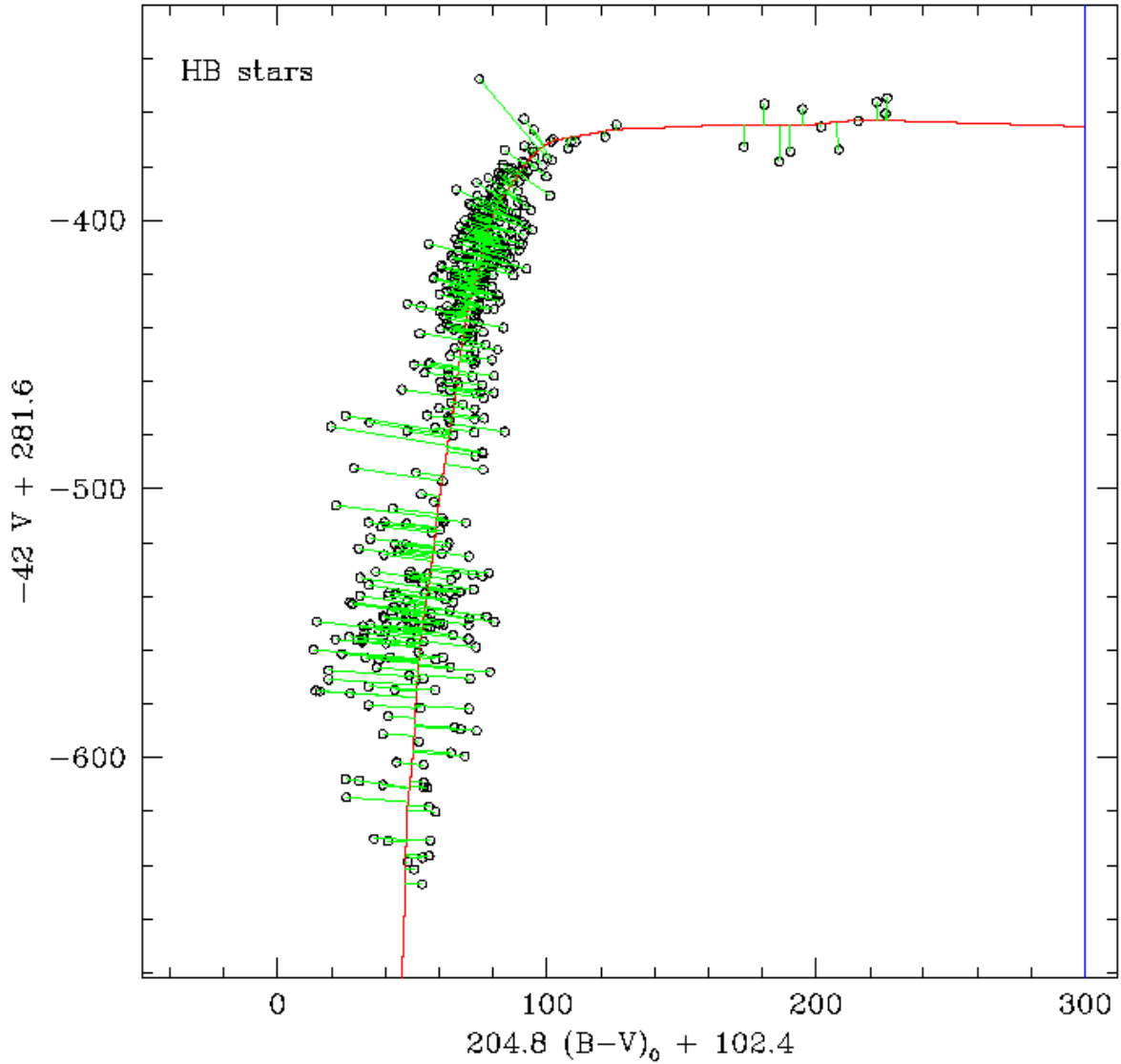,width=16cm}
\caption[projection.ps]{The diagram used to project the HB stars of
NGC~6273 onto
the HBRL. In this scale, 1 unit in the abscissa has the same length as 
1 unit in the ordinate. The HBRL and the projection vector of each star 
are shown. The vertical line on the right marks the adopted zero point
for $l_{\rm HB}$.
\label{pro}}
\end{figure}

\item For each star, the length $l$ of the HBRL from a fixed zero point of 
the coordinate, at $c=300$, to the point ($c_R$,$b_R$) was then computed
by dividing the HBRL into $N$ pieces of $\Delta c=0.01$, and approximating
each of these pieces by a straight line. 

\item Finally, the $l_{\rm HB}$ coordinate was obtained applying the
scaling relation:
\[l_{\rm HB}=0.1086 l\]
as in Dixon et al.\ (1996) for NGC~1904.

\end{itemize}

This procedure gives an $l_{\rm HB}$ on the same scale as in Ferraro
et al.\ (1992) and Dixon et al.\ (1996), with a different zero point 
for $l_{\rm HB}$, 
now set at $c=300$ (corresponding to $B-V=0.965$). The zero point
corresponds to a point redder than the reddest HB star in NGC~6441
(the most metal-rich of our clusters), in order to avoid negative
values for $l_{\rm HB}$. 


\subsection{The HB of NGC~6273}\label{HB19}

\begin{figure}
\psfig{figure=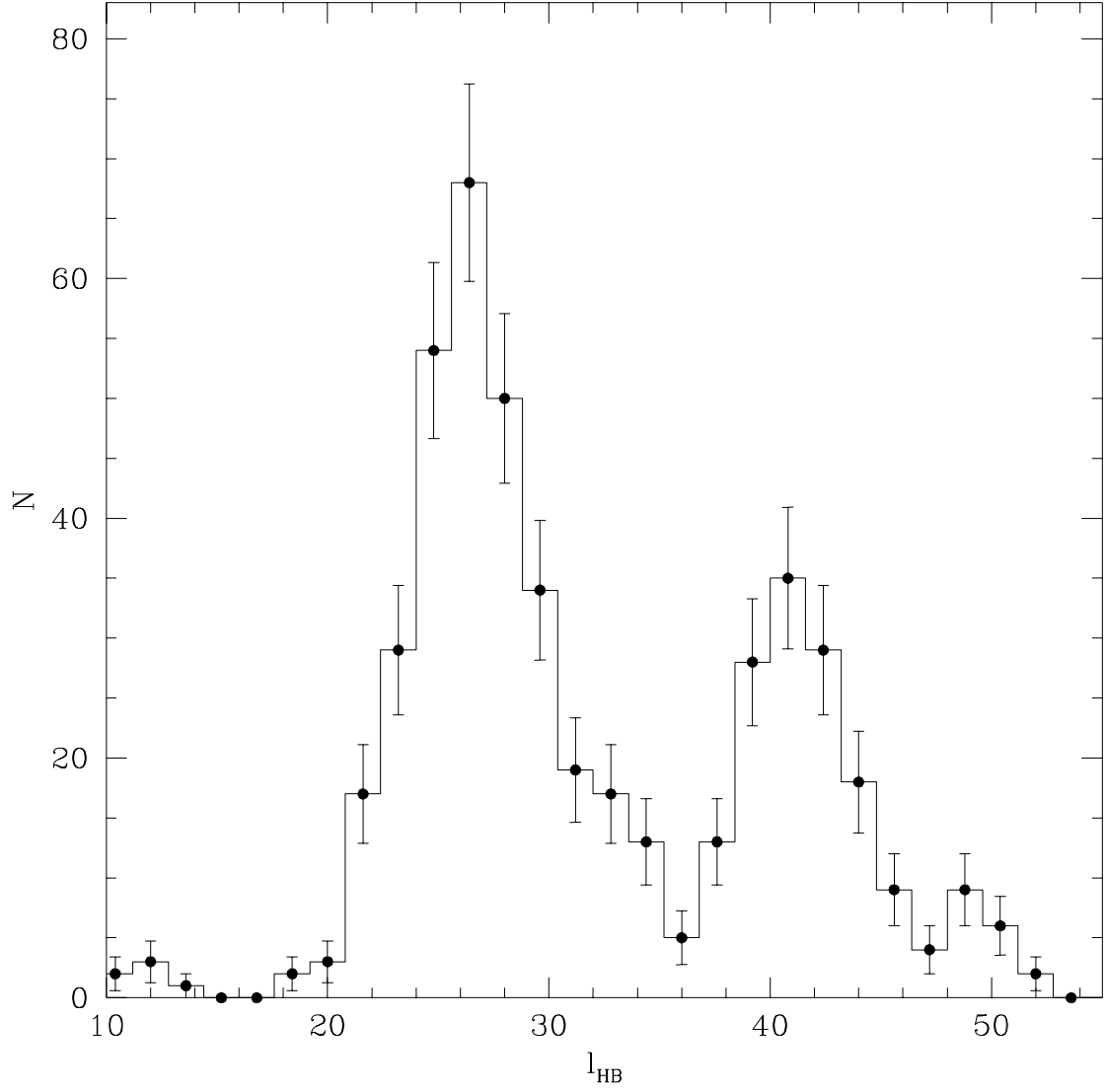,width=16cm}
\caption[histo.ps]{The distribution of the stars along the linear coordinate
on the HBRL is shown. The two peaks centered on $l_{\rm HB}=26.5$ and on 
$l_{\rm HB}=41.3$ and the large gap, centered on $l_{\rm HB}=36$, are clearly
visible.
\label{histo}}
\end{figure}

\begin{figure}
\psfig{figure=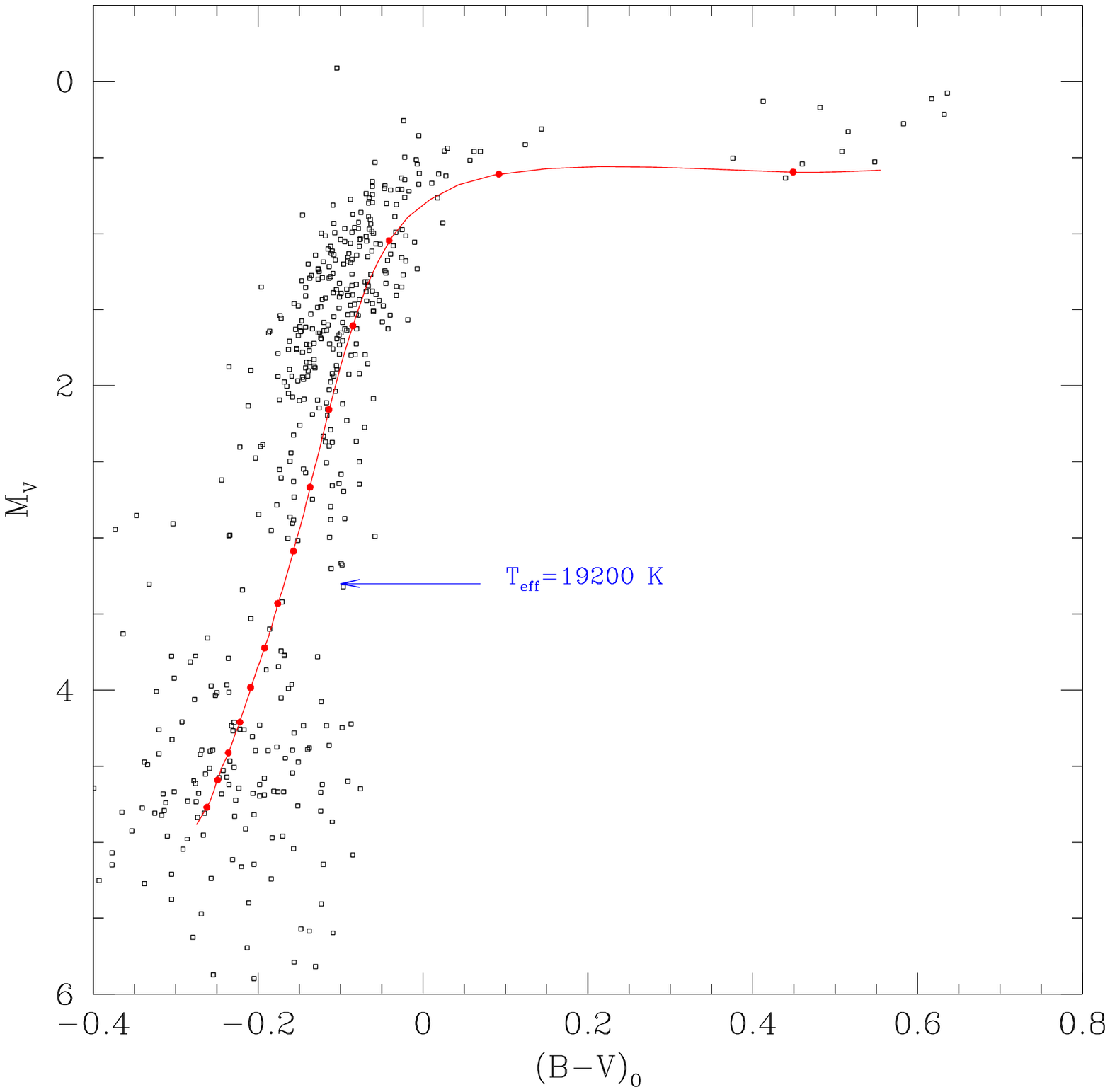,width=16cm}
\caption[modelhb.ps]{The HB stars of NGC~6273 with the model
by Bono et al.\ (1999) overplotted. The position and temperature of the
gap are indicated by the horizontal arrow. 
We adopted the reddening derived from the comparison with the BHB of
NGC~1904 (E($B-V$)=0.47) and applied a vertical shift of
$\Delta V=16.43$ in order to match the lower envelope of the star
distribution in the HB. In order to show the temperature scale, we marked with
full dots on the model the positions corresponding to temperatures from
6000 K (reddest dot) to 32,000K (bluest dot) in steps of 2000K.
\label{modelhb}}
\end{figure}

The $l_{\rm HB}$ distribution for NGC~6273 is shown in Fig.\
\ref{histo}.  The HB shows two remarkable peculiarities. First of all,
there is a blue tail which in the CMD extends well below the cluster turnoff.
Second, the EBT is clearly bimodal, with two peaks centered on $l_{\rm
HB}=26.5$ and on $l_{\rm HB}=41.3$ and a large gap, centered on $l_{\rm
HB}=36$. Apart from NGC~2808, none of the HBs with EBTs in our sample
(cf.  Section~\ref{otherhb}) shows such a clear bimodality in the
$l_{\rm HB}$ distribution.  Fig.\ \ref{histo} shows that the two peaks
are quite sharp.  It is questionable whether the distribution along the
EBT of NGC~6273 is peculiar because of the presence of these two peaks
or rather because of the gap. The two peaks can be fitted with 
gaussian functions centered on $l_{\rm HB}=26.5$ and on $l_{\rm
HB}=41.3$ with a sigma of 2.1 and 2 in $l_{\rm HB}$ units,
respectively. According to the new horizontal-branch models by Bono,
Cassisi, \& Castellani 
(1999), this dispersion in $l_{\rm HB}$ corresponds to a mass
dispersion $\Delta m = 0.017\> m_\odot$ on the HB.  Figure~\ref{modelhb}
shows an enlargment of the HB of NGC~6273 with overplotted the HB model
by Bono et al.\ (1999). The model has been fitted adopting 
the average reddening obtained in Section \ref{red} and allowing a 
vertical shift such that the model is the lower envelope of the
red HB stars.
According to these models, the gap in NGC~6273
corresponds to a $T_{\rm eff}\sim19,200$ K, and to a mass of $\sim0.54\>
m_\odot$.

\subsection{Multimodal HBs in other clusters}\label{otherhb}

Though the gap in Fig.\ \ref{histo} is remarkable, it is rather
difficult to assess its statistical significance. None of the methods
suggested by Catelan et al.\ (1998) can be applied to NGC~6273, as it
is not possible to know what is the underlying true $l_{\rm HB}$
distribution. Surely it cannot be uniform, as supposed by Ferraro et
al.\ (1998) for NGC~6093 and NGC~6205.  On the other hand, there are other
clusters which have blue tails and show gaps in the HB which might be
related to the gap in NGC~6273.  It is of some interest to check
whether these gaps are located in similar positions on the HBs of
different clusters. Ferraro et al.\ (1998) have shown that all the
clusters with EBTs have a gap on the lower part of the blue tail. They
have also suggested that at least one gap seems to be present at the
same location in all clusters, at $T_{\rm eff}\sim18,000$ K. 
The models in Fig.\ \ref{modelhb} seem to suggests a slightly hotter
temperature for the gap in NGC~6273. 
However, the absolute value of the temperature depends on the adopted
models (Ferraro et al.\ used the models by Dorman et al.\ 1997), and the
adopted transformation from the theoretical to the observational
plane. Moreover, the $V$ vs. $B-V$ plane is not the best one for
estimating the $T_{\rm eff}$ of such hot stars, which can be better
estimated with ultraviolet observations, as in Ferraro et al.\ (1997).

\begin{figure}
\psfig{figure=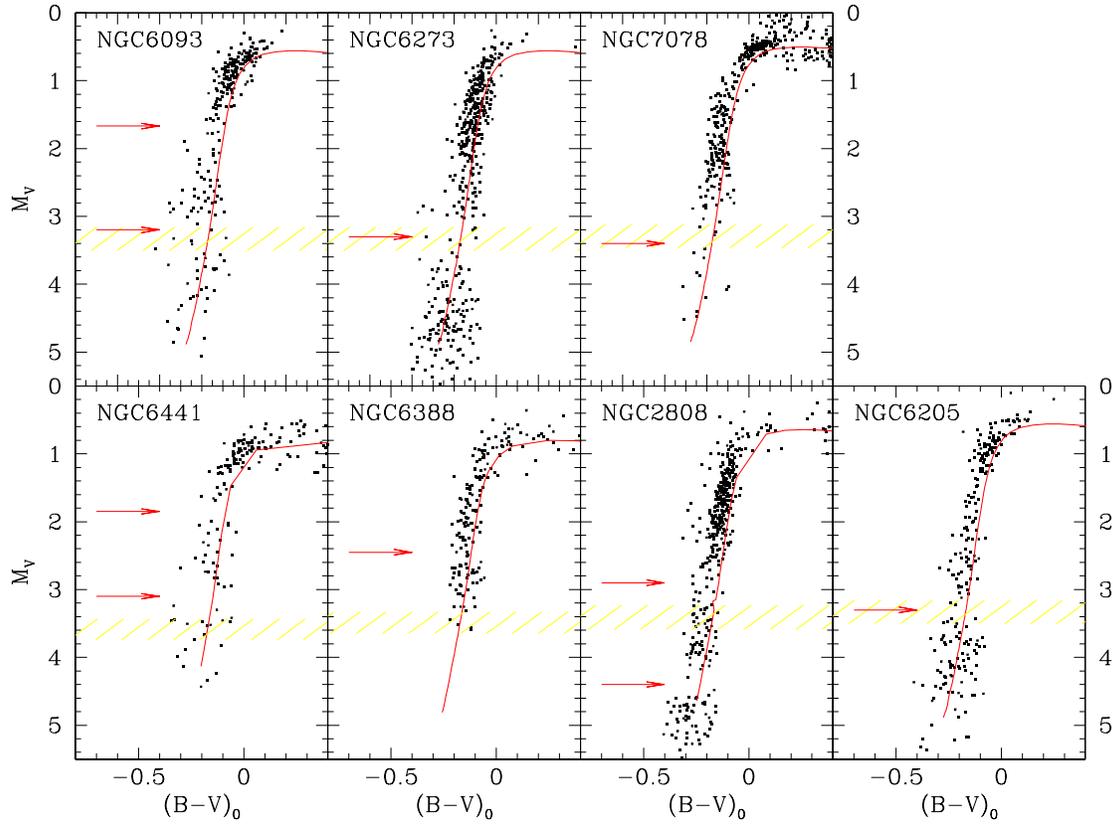,angle=-90,width=16cm}
\caption[cmd_Tcost.ps]{The ZAHBs of the seven clusters with gaps, 
matched with the models by Bono et al.\ (1999). Metallicity increases
from the lower-left box to the upper right.  The visible gaps are
marked with horizontal arrows. The shaded regions identify the
magnitude intervals that, according to the models, correspond to
$T_{{\rm eff}}=19,200\pm1000$K.
\label{cmd_Tcost}}
\end{figure}

In the following, we will try to take advantage of the fact that we
have eight clusters with blue HB tails observed exactly in the same
photometric system and with homogeneous photometry. We already noted
that our photometric bands are far from ideal for a study of the hot
population in the EBT of the GGCs, and that, due to the saturation of
the ($B-V$) color for stars hotter than $\sim 10,000$K the HB becomes
almost vertical, making more difficult the identification of all the
possible gaps (real or statistical fluctuations as they might be).
A typical example is NGC~1904, for which Hill et al.\ (1996) suggested
a possible gap at $T_{\rm eff}\sim
9,990$K, which is not visible in our HST CMD (Sosin et al.\ 1997b).
Still, in Figs.\ \ref{cmd_Tcost} and \ref{histo_Tcost}, there
is clear evidence of gaps (the most significant ones are indicated by
arrows), so we can still try to address the question of whether
there is anything systematic about their positions.

One possible way to compare the location of the gaps in different
clusters is to compare their $l_{\rm HB}$ directly. This method has
the advantage of being totally empirical. However, in turn it makes
very hard the interpretation of the different locations of the gaps in
terms of physical parameters like $T_{\rm eff}$ or mass.  This is
mainly due to the fact that, by definition, $l_{\rm HB}$ does not run
over the ZAHB, but over the mean HB line; therefore the comparison
between observed data and models (which refer to the ZAHB) is not
completely consistent.  On the other hand, forcing $l_{\rm HB}$ to run
over the ZAHB (\ie, the model) instead of on the mean ridge line, may
introduce strong biases in the projections of the stars in those
regions where the shape of the model does not reproduce exactly the
observed HB. Finally, the models often do not reach magnitudes as
faint as the data do, in the EBT. For these reasons, and since all the
gaps visible in our CMDs are actually located in the vertical part of
the HB, we decided that a more direct analysis could be done by simply
comparing the distribution in absolute magnitude of the vertical part
of the HBs of the different clusters.

\begin{figure}
\psfig{figure=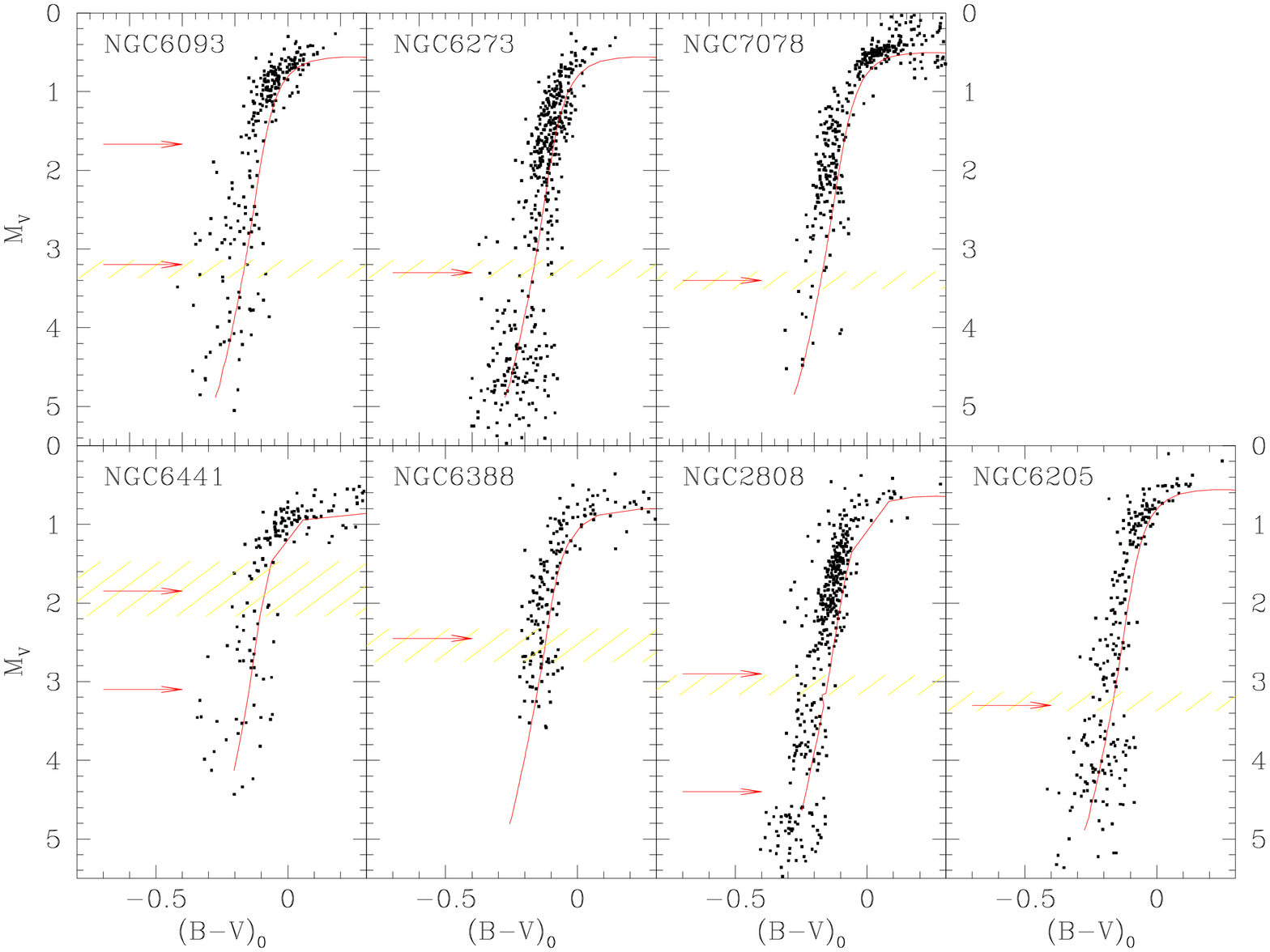,angle=-90,width=16cm}
\caption[cmd_Mcost.ps]{Same as Fig.\ \ref{cmd_Tcost} but with the
shaded region identifying the region corresponding to the interval
$0.53<M/M_\odot<0.54$.
\label{cmd_Mcost}}
\end{figure}

To this end, we proceeded as follows. We registered the blue bend of
all the HBs to that of NGC~6205, in order to constrain the relative
reddening. We then assumed an $E(B-V)=0.02$ for NGC~6205 and
de-reddened accordingly the HBs of all the other clusters.  Allowing
only a vertical shift we then matched the lower envelope of the stars
in the observed HBs with the models by Bono et al.\ (1999), as shown by
Fig.\ \ref{cmd_Tcost}. In this figure, the most prominent HB gaps are
indicated with the horizontal arrows. Following the suggestion by
Ferraro et al.\ (1998) we marked with a dashed horizontal strip the
region corresponding to $T_{\rm eff}=19,200\pm1000$K (i.e., the
temperature at the center of the gap of NGC~6273).  At least one gap
of the intermediate-metallicity and metal-poor clusters is always
inside this strip, confirming their hypothesis that in all the EBTs
there is a gap corresponding to this temperature. 
However, for the more metal-rich clusters, namely NGC~6441, NGC~6388,
and NGC~2808, there is no gap corresponding to the dashed areas in
Fig~\ref{cmd_Tcost}. While in the case of NGC~6388 the region at
$T_{\rm eff}=19,200\pm1000$K is in any case in a sparsely populated
part of the CMD, in NGC~6441 and NGC~2808 it corresponds to local peaks in 
the distribution along the HB.
Note that in the case of NGC~2808, all the gaps visible in the
ultraviolet regions ($B$ vs. F214W$-B$ diagram) are also visible in
the $V$ vs. $B-V$ CMD (Sosin et al.\ 1997). We do not have suitable UV data for
NGC~6388 and NGC~6441.
Shall we conclude that
gaps are at random positions in the HB? Of course, the uncertainties in
the models and in the transformation from the theoretical to the
observational plane, and the errors in fitting the models to the
observed HB, play an important role in the interpretation of
Fig.\ \ref{cmd_Tcost}; we cannot exclude that the hypothesis in Ferraro
et al.\ (1998) is still valid.  On the other hand, it is possible that
some of the gaps marked in Figs.\ \ref{cmd_Tcost} and \ref{histo_Tcost}
are a random fluctuation in the distribution of the stars along the HB
(Catelan et al.\ 1998).  We have also tested an alternative hypothesis.
Since the main parameter governing the position of a star along the HB
is the ratio of the envelope mass to the core mass (a ratio that 
follows from the total mass, assuming that the core mass is the
same for all these stars), we have investigated the possibility that
the gaps appear at costant mass on the HB. Fig.\
\ref{cmd_Mcost} shows the same HBs of Fig.\ \ref{cmd_Tcost} with a
shaded region indicating the magnitude corresponding to
$M=0.535\pm0.005M_\odot$. At least one gap of each cluster falls
inside this area, suggesting that the possibility of the gaps being
physically associated with a ``forbidden'' value of the mass is
compatible with the observations.  The analysis of the histograms, in
absolute magnitude, of the stars in the vertical portion of the HBs
gives the same result.  Figures~\ref{histo_Tcost} and
\ref{histo_Mcost} show the distribution of the same stars plotted in
Figs.\ \ref{cmd_Tcost} and \ref{cmd_Mcost} as a function of
magnitude. Again, the shaded
region indicates the constant-temperature loci (Fig.\
\ref{histo_Tcost}) and the constant-mass loci (Fig.\
\ref{histo_Mcost}).  
And again, the hypothesis of a set of gaps at constant mass for all the
clusters with EBT is in agreement with all the data at our disposal.

In conclusion, the distribution of the stars along the blue HB tails
is clearly not uniform and differs from cluster to cluster. There are
both peaks and gaps.  All the clusters with an EBT have at least one
gap, as suggested by Ferraro et al.\ (1998).  Of course a sample of
only seven clusters is not sufficient to draw any conclusion at this
point. Moreover, as mentioned above, the $V,B-V$ plane is not the
ideal one for these studies.  Still,
Figures~\ref{cmd_Tcost}--\ref{histo_Mcost}
suggest that the possibility that gaps are at constant mass is at
least as plausible as the hypothesis of gaps at constant temperature as
suggested by Ferraro et al.\ (1998).  There are also other gaps, which
apparently are not directly related to either of the two proposed
hypotheses. 

\begin{figure}
\psfig{figure=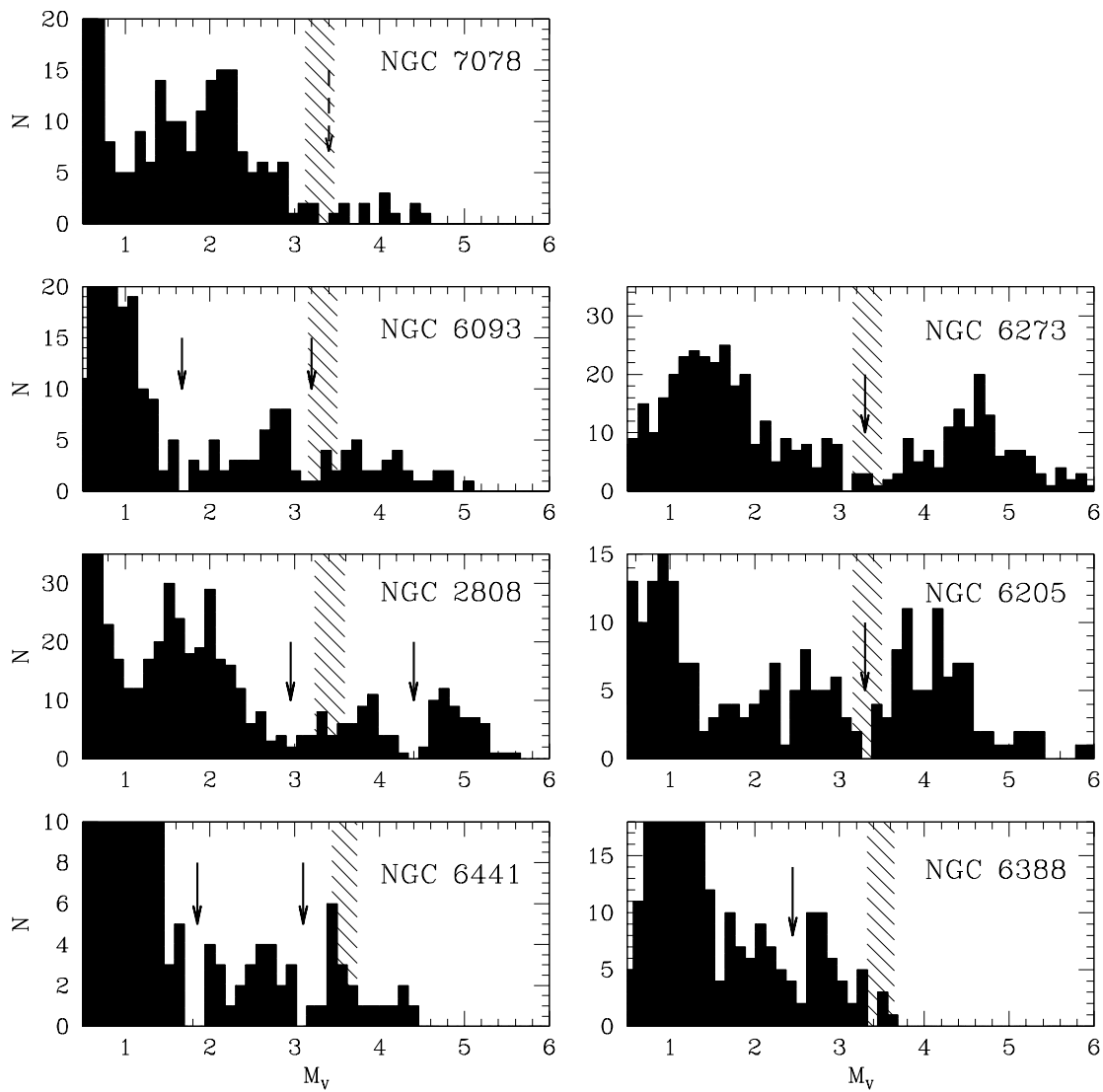,angle=0,width=16cm}
\caption[cmd_Tcost.ps]{Distribution in magnitude of the stars
in the vertical region of the HBs shown in Figs.\ \ref{cmd_Tcost} and
\ref{cmd_Mcost}. Metallicity decreases from the lower-left panel to
the top. The shaded regions identify the magnitudes intervals that,
according to the models, correspond to $T_{{\rm eff}}=19,200\pm1000$K
\label{histo_Tcost}}
\end{figure}

\begin{figure}
\psfig{figure=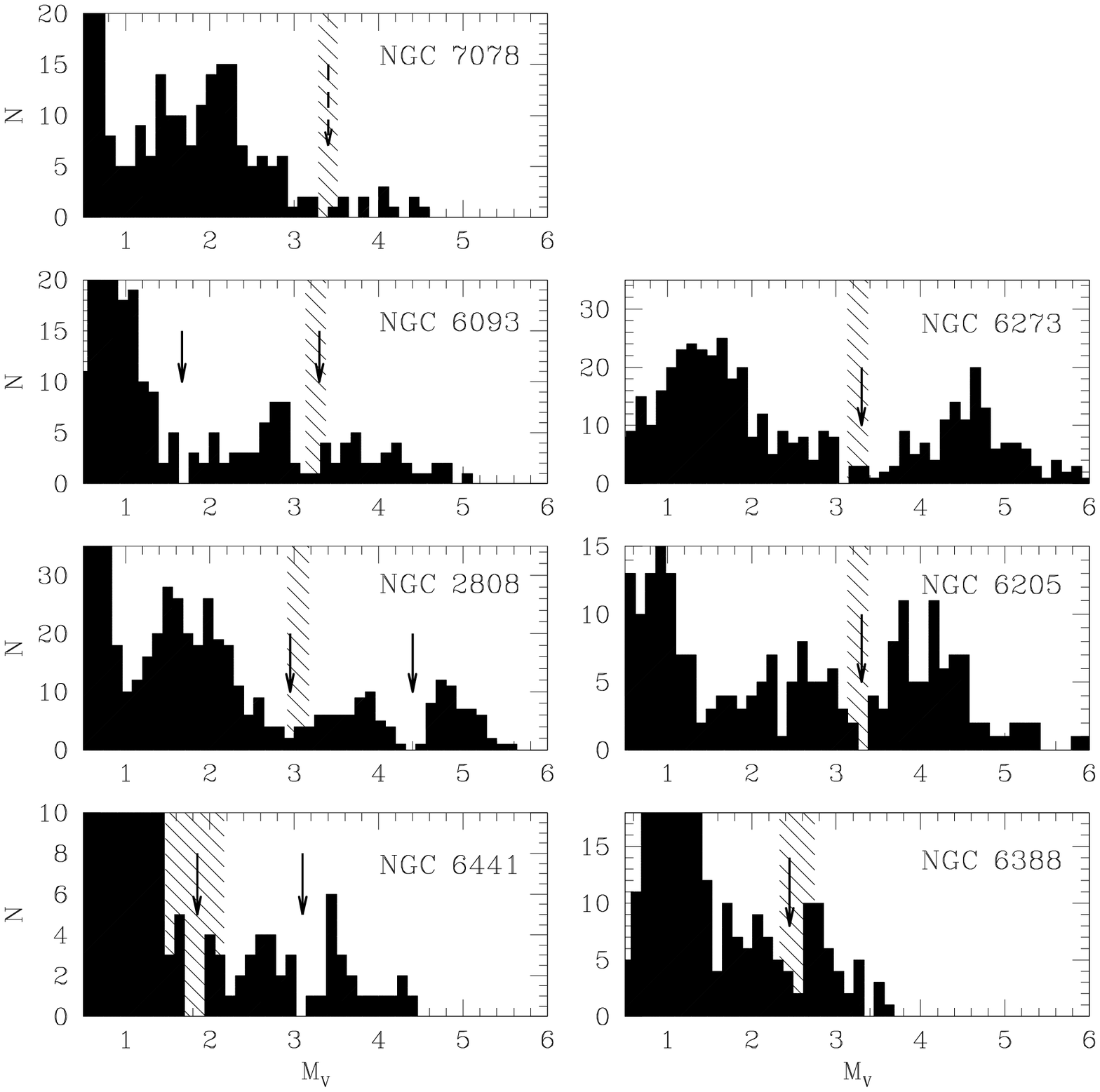,angle=0,width=16cm}
\caption[cmd_Mcost.ps]{Same as Fig.\ \ref{histo_Tcost} but with the
shaded region identifying the magnitudes corresponding to the interval
$0.53<M/M_\odot<0.54$.
\label{histo_Mcost}}
\end{figure}


\section{Luminosity Functions}\label{lf}

>From the CMD of Fig.\ \ref{cmdco} it is possible to extract a LF for
the evolved stars, which is an important direct test of the evolutionary 
clock (Renzini \& Fusi Pecci 1988).
It has been recently pointed out that, at least for the metal-poor
clusters, the agreement of the models with the observed LFs for the
GGC stars is far from satisfactory (Faulkner \& Swenson 1993, Bolte 1994,
and references therein).  Sandquist et al.\ (1998), by comparing the LF
of M30 with the theoretical LFs from Bergbusch \& VandenBerg (1992) 
and VandenBerg et al.\ (1999), confirm this result, at least for
the most metal-poor clusters, and suggest that it might be the
consequence of some deep mixing events. On the other hand, Silvestri
et al.\ (1998) claim that their set of models strongly reduces
the discrepancy with the observed LFs, though they cannot provide the
reasons for the differences among the different models.

The rich fields in the centers of the GGCs whose CMDs we are
collecting are suitable for a detailed study of the
evolution time scales along the red giant branch. In Piotto et al.\
(1999a) we already presented a comparison between the Bergbusch \&
VandenBerg (1992) LFs and the observed LFs for NGC~6362 and NGC~6934,
with conflicting results.  While the LF of NGC~6934 ([Fe/H] $=-1.03$)
could be reasonably well fitted with the theoretical LFs by Bergbusch
\& VandenBerg (1992), NGC~6362 ([Fe/H] $=-1.48$) showed the same
excess of bright giants as suggested by Bolte (1994) for the most
metal-poor clusters.

We first tried the same fit as in Piotto et al.\ (1999a), using the 
appropriate set of isochrones from Bergbusch \& VandenBerg (1992), on
the observed LF of NGC~6273, after the crowding corrections. Again, we
found an overabundance of bright RGB stars with respect to the MS,
as for NGC~6362 or M30 (Sandquist et al.\ 1998).

\begin{figure}
\psfig{figure=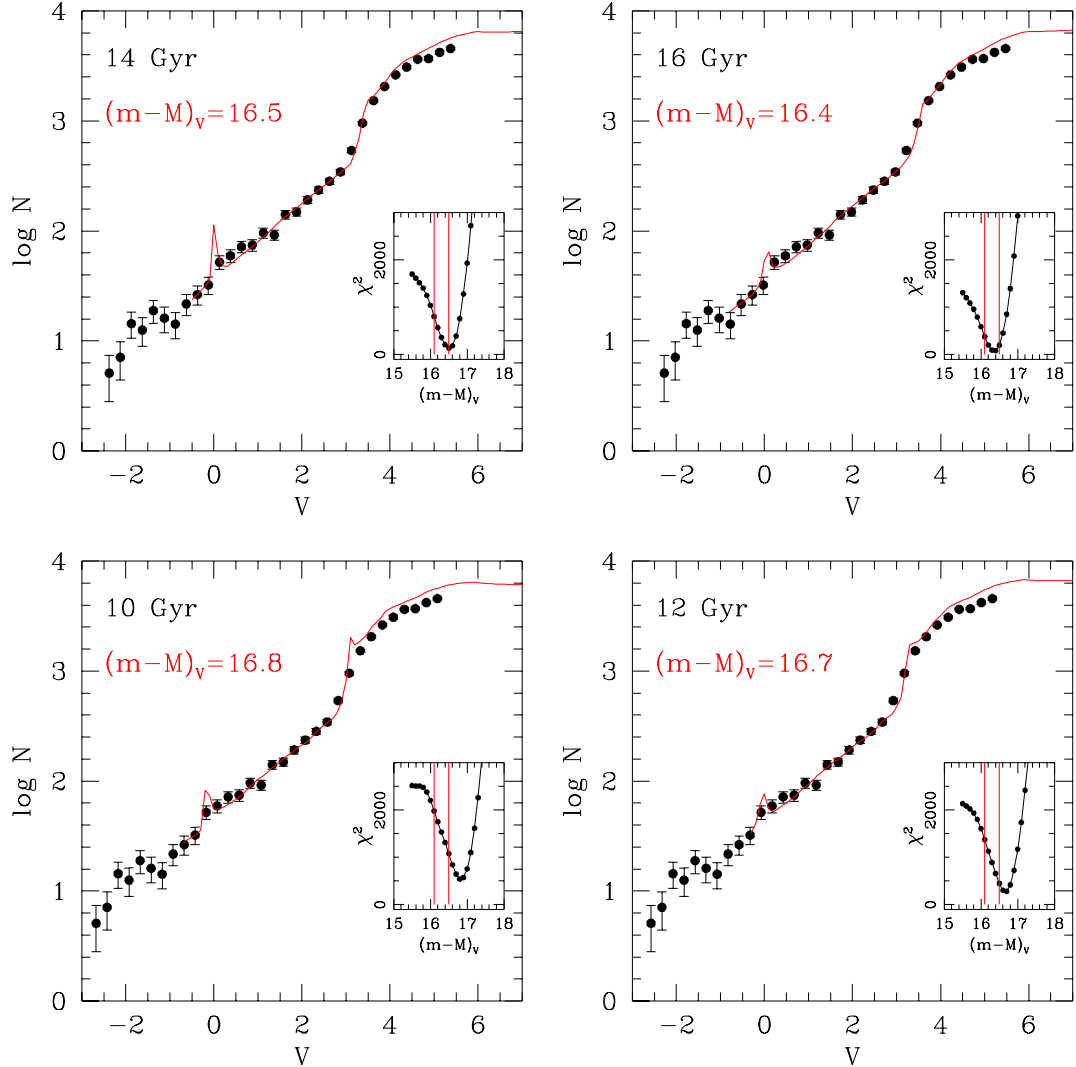,width=16cm}
\caption[lfdan.ps]{The RGB LF of NGC~6273 ({\it solid dots}) is compared
with the models by Silvestri et al.\ (1998) ({\it solid lines}) for
$Z=0.0003$.  The four different windows show different age models. The
behaviour of the $\chi^2$ as a function of different distances is shown
in the small panel inside each window:\ the minimum corresponds to the
$(m-M)_V$ required for that particular fit (this value is given in
the label). The two vertical lines in the $\chi^2$ panels are the limits
($\pm$ 0.2 mags around the value determined in Section~\ref{dist})
outside which we consider that the required distance is not realistic.
\label{lfdan}}
\end{figure}

\begin{figure}
\psfig{figure=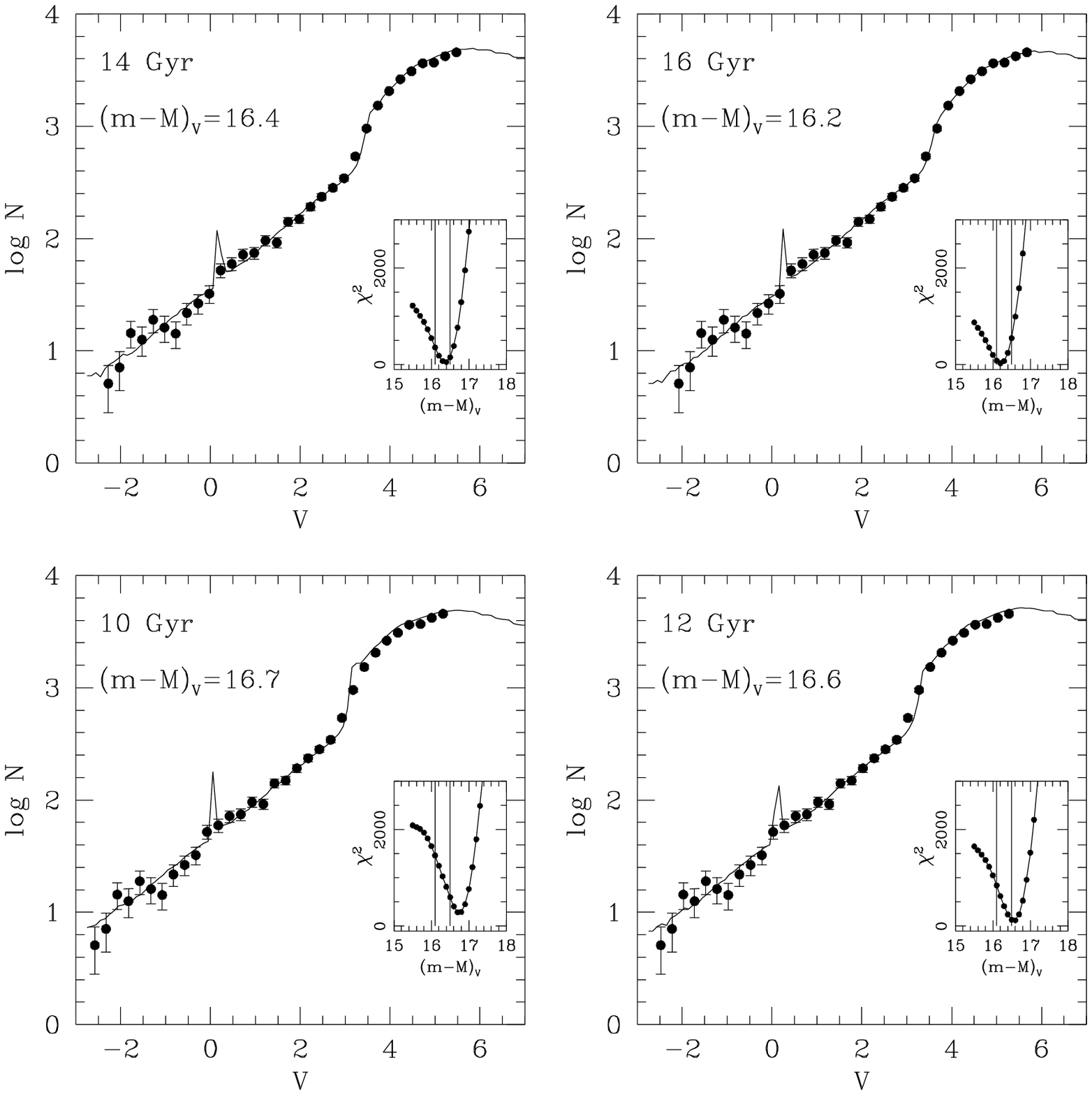,width=16cm}
\caption[lfchi.ps]{Same as Fig.\ \ref{lfdan}, but for the models by
Straniero et al.\ (1997) and $Z=0.0004$.
\label{lfchi}}
\end{figure}

A better fit could be obtained using both the most recent models by
Silvestri et al.\ (1998), and the models by Straniero, Chieffi \&
Limongi (1997). In order to mimic an $\alpha$-element enhancement of
0.4 (Sneden et al.\ 1991), we adopted a higher metallicity than
discussed in Section~\ref{intro}, following the method described by Salaris et
al.\ (1993).  We also adopted a flat mass function ($x=-0.5$, where the
Salpeter slope is 1.35), which is more appropriate for the GC center. 
In both cases there is a good agreement between the
observed and theoretical LFs (Fig.\ \ref{lfdan} and Fig.\ \ref{lfchi}),
which mostly removes the claimed discrepancies with the calculated
evolutionary times along the RGB. We note that the models by Straniero
et al.\ (1997) seem to reproduce the observed stellar distribution better.

Piotto et al.\ (1999a) also suggested that the large number of evolved
stars in the LF from the cluster cores could be used to infer
the cluster ages independently. Silvestri et al.\ (1998) pointed out that the
method can work, provided we have an independent distance
estimate. This is also evident from Figs.\ \ref{lfdan} and
\ref{lfchi}. Each panel shows the fit of the NGC~6273 LF with the
theoretical LFs for 4 different ages.  The inset in each panel shows
the $\chi^2$ trend as a function of the distance.  The minimum value
of the $\chi^2$ is similar in the three panels corresponding to an age 
$\geq12$ Gyr, while it is higher for 10 Gyr. Note that if we allow
the distance to vary, there is no preferred age between 12 and 16 Gyr.
However, this would imply an uncertainty in the apparent distance
modulus at a level of more than 0.4 magnitude, which is not
realistic.  On the other hand, if we have some estimate of the
distance, the uncertainty in the age is drastically reduced.  For
example, if we take the distance estimate from Section~\ref{dist} we
are forced to conclude that the age of NGC~6273 is $15\pm2$ Gyr (from
the Straniero et al.\ 1997 models, slightly older than from the Silvestri et
al.\ models). Note that the distance obtained in Section~\ref{dist} is
still based on the old, shorter distance scale. If we adopt the longer
distance scale from the HIPPARCOS data, as in Gratton et al.\ (1997), we
should reduce the ages by 2 Gyr.

It remains still to verify the suggestion by Bergbusch \& VandenBerg
(1992) that the detailed structure of the LF around the TO can constrain
both the age and the distance. The smearing of the LF due to the
differential reddening does not allow applying the method to NGC~6273,
however.

\section{Acknowledgments}\label{ack}

\acknowledgments

We thank P.\ Stetson for his generosity with software.  It is a pleasure
to thank M.\ Catelan for useful discussions. 
We are indebted to S. Cassisi for providing us a set of HB models
specifically calculated for the metallicities of our clusters.
Support for this work was provided to I.R.K., S.G.D., and R.M.R.\ by
NASA through grant GO-7470 from the Space Telescope Science Institute.
G.P.\ and M.Z.\ acknowledge partial support by the Ministero
dell'Universit\`a e della Ricerca Scientifica.
\newpage



\begin{references}

\reference{} Bolte, M. 1994, \apj, 431, 223
\reference{} Bono, G., Cassisi, S., \& Castellani, V. 1999, in prepation
\reference{} Bergbusch, P.~A., \&\ VandenBerg, D.~A. 1992, \apjs, 81, 163 
\reference{} Catelan, M., Borissova, J., Sweigart, A.~V., Spassova, N.,
	1998, \apj, 494, 265
\reference{} Clement, C., \& Sawyer Hogg, H., 1978, \aj, 83, 167
\reference{} Cool, A.~M., Piotto, G., \&\ King, I.~R. 1996, \apj, 468, 655
\reference{} Dixon, W.~V.~D., Davidsen, A.~F., Dorman, B., \& Ferguson,
	H.~C., 1996, \aj, 111, 1936
\reference{} Djorgovski, S. 1993, in Structure and Dynamics of Globular
	Clusters, eds.\ S.\ G.\ Djorgovski \& G.\ Meylan (San 
	Francisco:\ ASP),p.\ 373
\reference{} Ferraro, F.~R., Clementini, G., Fusi Pecci, F., Sortino, R.,
	\& Buonanno, R. 1992, \mnras, 256, 391
\reference{} Ferraro, F.~R., et al.\ 1997, \aa, 324, 915
\reference{} Ferraro, F.~R., Paltrinieri, B., Fusi Pecci, F., Rood, R.~T.,
	\& Dorman, B., 1998, \apj, 311, 319
\reference{} Faulkner, J., \& Swenson, F.~J. 1993, \apj, 411, 200
\reference{} Fusi Pecci, F., Ferraro, F.~R., Bellazzini, M., Djorgovski, 
	S.~G., Piotto, G., \& Buonanno, R., 1993, \aj, 105, 1145
\reference{} Gratton R.~E., Fusi Pecci, F., Carretta, E., Clementini, G.,
	Corsi, C., \& Lattanzi, M. 1997, \apj, 491, 749
\reference{} Harris, W.~E., Racine, R., \& de Roux, J., 1976, \apjs, 31, 13
\reference{} Hill, R.~S., 1996, \aj, 112, 601
\reference{} Holtzman, J.~A., Burrows, C.~J., Casertano, S., Hester, J.~J., 
	Watson, A.~M., \& Worthey, G.~S. 1995, PASP, 107, 1065
\reference{} Iben, I. 1968, Nature, 220, 143
\reference{} Norris, J., 1983, \apj, 272, 245
\reference{} Piotto, G., Zoccali, M., King, I.~R., Djorgovski, S.~G.,
	Sosin, C., Dorman, B., \& Rich, R.~M. 1999a, \aj, 117, 264
\reference{} Piotto, G., et al. 1999b, in preparation
\reference{} Renzini, A., \& Fusi Pecci, F. 1988,  ARA\&A, 26, 199
\reference{} Rood, R.~T., \& Crocker, D.~A., 1985, in Horizontal Branch
	and UV-Bright Stars, ed.~A.~G.~D.\ Philip (Schenectady: L.\ Davis 
	Press), 99
\reference{} Rood, R.~T., \& Crocker, D.~A. 1989, in The
	Use of Pulsating Stars in Fundamental Problems of Astronomy (IAU
	Colloq.~111), ed.\ E.\ G.\ Schmidt (Cambridge:\ Cambridge Univ.\ 
	Press),p.\ 103
\reference{} Rutledge, G.~A., Hesser, J.~E., Stetson, P.~B., Mateo, M.,
	Simard, L., Bolte, M., Friel, E.~D., \& Copin, Y., 1997, PASP, 
	109, 883
\reference{} Salaris, M., Chieffi, A.,\& Straniero, O., 1993, \apj, 
	414, 580
\reference{} Sandquist, E.~L., Bolte, M., \& Hernquist, L. 1997, \apj, 477, 
	335
\reference{} Silbermann, N.~A., et al.\ 1996, \apj, 470, 1
\reference{} Silvestri, F., Ventura, P., D'Antona, F., \& Mazzitelli, I., 
	1998, \apj, submitted
\reference{} Sneden, C., Kraft, R.~P., Prosser, C.~F., \& Langer, G.~E., 1991,
	\aj, 102, 2001
\reference{} Sosin, C., Dorman, B., Djorgovski, S.~G., Piotto, G., Rich,
	R.~M., King, I.~R., Liebert, J., Phinney, E.~S., \& Renzini, A. 
	1997a\apj, 480, L35
\reference{} Sosin, C., Piotto, G., King, I.~R., Djorgovski, S.~G., 
	Rich, R.~M., King, I.~R., Dorman, B., Liebert, J., \& Renzini, 
	A. 1997b in Advances in Stellar Evolution, eds.\ R.\ T.\ Rood 
	and A.\ Renzini (Cambridge:\ Cambridge Univ.\ Press), p. 92
\reference{} Stetson, P.~B., \& Harris, W.~E., 1977, \aj, 82, 954
\reference{} Straniero, O., Chieffi, A., \& Limongi, M., 1997, \apj, 490, 425
\reference{} VandenBerg D.~A., Swenson F.~J., Rogers F.~J., Iglesias C.~A., 
	\& Alexander D.~R., 1998, \apj, submitted
\reference{} White, R.~E., \& Shawl, S.~J., 1987, \apj, 317, 246
\reference{} Zinn, R., \& West, M. 1984, \apjs, 55, 45
\reference{} Zoccali, M., \& Piotto, G., 1999, in preparation

\end{references}
\end{document}